\documentstyle[preprint,aps]{revtex}
\title{\bf Benchmark calculation of the 
three-nucleon photodisintegration}

\author{J. Golak$^{1,2}$, R. Skibi\'nski$^{2}$, W. Gl\"ockle$^{1}$, 
H. Kamada$^{3}$, A. Nogga$^{4}$,  H. Wita{\l}a$^{2}$}
\address{
$^1$ Institut f\"ur Theoretische Physik II, Ruhr-Universit\"at 
Bochum, D-44780 Bochum, Germany\\
$^2$ Institute of Physics, Jagellonian University, PL-30059 Cracow,
Poland\\
$^3$ Department of Physics, Faculty of Engineering, Kyushu Institute 
of Technology \\ 1-1 Sensucho, Tobata, Kitakyushu 804-8550, Japan\\
$^4$ Department of Physics, University of Arizona, Tucson, Arizona 85721,
USA}

\author{
V. D. Efros$^{5}$, W. Leidemann$^{6}$, G. 
Orlandini$^{6}$ and E.L. Tomusiak$^{7}$}
\address{$^5$ Russian Research Centre "Kurchatov Institute", Kurchatov 
Square 1, 123182 Moscow, Russia\\
$^6$ Dipartimento di Fisica, Universit\`a di Trento and INFN (Gruppo collegato
di Trento) \\ I-38050 Povo (Trento), Italy\\ 
$^7$ Department of Physics and Astronomy, University of Victoria,
Victoria BC V8P 1A1, Canada  
}

\begin{document}

\date{\today}

\maketitle
\begin{abstract}
A benchmark is set on the three--nucleon  
photodisintegration calculating the total 
cross section with modern realistic two-- and three--nucleon 
forces (AV18, UrbIX) using both the Faddeev equations and the 
Lorentz Integral Transform method. 
This test shows that the precision of three-body calculations 
involving continuum states is considerably higher than experimental
uncertainties. 
Effects due to retardations, higher multipoles, 
meson exchange currents and Coulomb force are studied. 
\end{abstract}

\noindent {\footnotesize 
}

\vfill\eject
\section{INTRODUCTION}

The total photodisintegration cross sections of the three--body
nuclei are important quantities for understanding electromagnetic processes 
in nuclei. 
Several approximate theoretical calculations have been 
performed in the past with simple 
NN-potential models \cite{BaP70,GiL76,Fang78,Vost81}. 
The first calculations with complete final state interaction 
were carried out with the Lorentz Integral 
Transform (LIT) method using semirealistic \cite{ELO97}
and realistic NN and 3N forces \cite{ELOT00,ELOT01}. 
In this paper we intend to set a benchmark on the results 
of the three--nucleon total photodisintegration by calculating the  
cross section with modern realistic two-- and three--nucleon 
forces. To this end we use two different methods, namely the 
Faddeev and the LIT approaches. 
On the one hand the comparison between the results gives an idea 
of the precision reached 
by three-body calculations involving continuum states while on the 
other hand it may stimulate more accurate measurements.

It has long been believed that the photonuclear cross 
section is dominated by unretarded dipole transitions.
The contributions of magnetic and higher order electric multipoles,
of retardation, 
as well as of meson exchange currents 
have been discussed extensively. However, rigorous results 
have only been given for the deuteron \cite{Ar74,Bu85,Ar91}. 
In this paper we give for the first time a quantitative
discussion of these issues for the three-body nuclei.

The paper is organized as follows.
In Sec. 2 we briefly summarize the two methods for obtaining
the total photodisintegration cross section. In Sec.
3 we present the comparison between the two calculations 
and discuss the effects of the Coulomb force, as well as  
the contributions beyond the unretarded dipole operator. 
Conclusions are drawn at the end.

\section{THEORETICAL FRAMEWORK}
%************************   end  ******************************

The total unpolarized photodisintegration cross section 
is given by 
\begin{equation}\label{sigmat}
\sigma_T (E_\gamma) = R_T(E_\gamma) \, \frac1{E_\gamma} 4\pi^2 
\frac{e^2}{\hbar c}\,.
\label{eqfad0}
\end{equation}
In terms of the matrix elements of the transverse current operator 
${\bf J_T}$ the transverse response function $R_T(E_\gamma)$ is

\begin{equation}\label{rt}
R_T(E_{\gamma})=\int df |\langle\Psi_f|{\bf J_T}|\Psi_0\rangle|^2
              \delta(E_{\gamma}-E_f+E_0-\frac{E_\gamma^2}{2 M_t})\,,  
\end{equation}
where $\Psi_0$ is the three--nucleon bound state wave function with  
energy $E_0$, and $\Psi_f$ are final state wave functions. The term 
$E_\gamma^2/(2M_t)$ in the energy conserving delta--function 
represents the small recoil energy of the whole 3--body nucleus
with mass $M_t$.
%**************************   begin   *******************************
In the following subsections it is described how $R_T$ is 
calculated using the Faddeev equations and the LIT method. 

\subsection{The Faddeev equations}
%\section{The Faddeev equations}

%**************************   end   *******************************
\noindent
The way the transverse response function $R_T$ is calculated in the
Faddeev framework for inclusive electron scattering without a 3N 
force is described in Refs. \cite{Gol95} and \cite{Ishi98}. Here 
we sketch the corresponding derivation for the case when a 3N force
is additionally taken into account.

\noindent
 For a general operator ${\cal O}$ and using closure the function
in Eq.~(\ref{rt}) is rewritten as 
\begin{equation}
R_{\cal O} ( E_\gamma ) = - \frac1{\pi} \, \Im \, \langle \Psi_0 | 
{\cal O}^\dagger \frac1{E + i \epsilon - H} \, {\cal O} 
| \Psi_0 \rangle \,,
\label{eqfad2}
\end{equation}
where $E=E_0-E_\gamma^2/(2M_t)+E_{\gamma}$ and $H$ is the 3N 
Hamiltonian.
Next we define an auxiliary state 
\begin{equation} 
| \Psi \rangle \equiv 
\frac1{E + i \epsilon - H} \, {\cal O} | \Psi_0 \rangle\, , 
\label{eqfad3}
\end{equation}
which fulfils the following equation 
\begin{equation}\label{lorentzfad} 
(E + i \epsilon - H) \, | \Psi \rangle = {\cal O}| \Psi_0 \rangle\,.
\label{eqfad4}
\end{equation}
This leads to an intermediate equation
\begin{equation}
| \Psi \rangle \ = \
G_0 \, ( V_1 + V_2 + V_3 + V_4 ) | \Psi \rangle  \ + \ 
G_0 \, {\cal O} | \Psi_0 \rangle .
\label{eqfad5}
\end{equation}
Here $V_i$ is the NN potential (we use the notation $V_1 = V_{23}$ 
etc.), $V_4$ is a 3N force and $G_0$ the free 3N propagator.
Both the operator $\cal O$ and the 3N force $V_4$ can be written
as a sum of three parts having the same symmetry under particle 
exchanges.
\begin{eqnarray}
{\cal O}= \sum_{i=1}^3 {\cal O}_i \,, \\ \nonumber
V_4 = \sum_{i=1}^3 V_4^{(i)}
%\label{eqfad6}
\end{eqnarray}
Introducing the Faddeev decomposition of the state $| \Psi \rangle $ 
\begin{equation}
| \Psi \rangle \ = \ G_0 \, \sum_{i=1}^3 | U_i \rangle
\label{eqfad7}
\end{equation}
we obtain from Eq.~(\ref{eqfad5}) (for three identical nucleons) 
the following equation on $| U_1 \rangle$:
\begin{equation}
( 1 - V_1 G_0) \, | U_1 \rangle \ = \
V_1 G_0 P \, | U_1 \rangle  \ + \
V_4^{(1)} G_0 ( 1 + P ) \, | U_1 \rangle  \ + \
{\cal O}_1 | \Psi_0 \rangle ,
\label{eqfad8}
\end{equation}
where $P$ is the sum of a cyclical and
anticyclical permutation of 3 particles.
Using the identities for the NN t-operator $t_1$
\begin{eqnarray}
( 1 - V_1 G_0)^{-1} \ = \ 1 + t_1 G_0 \,, \\ \nonumber
( 1 - V_1 G_0)^{-1} \, V_1 G_0 \ = \ t_1 G_0
%\label{eqfad9}
\end{eqnarray}
we obtain the final Faddeev-like equation on $| U_1 \rangle$:
\begin{equation}\label{U1}
| U_1 \rangle \ = \
( 1 + t_1 G_0 ) \, {\cal O}_1 | \Psi_0 \rangle \ + \
\left( t_1 G_0 P \ + \ ( 1 + t_1 G_0 ) \, V_4^{(1)} ( 1 + P ) G_0 
\right)  \, | U_1 \rangle  .
\label{eqfad10}
\end{equation}
The response function $R_T$ is then obtained as 
\begin{equation}
R_T( E_\gamma ) = - \frac3{\pi} \, \Im \, \langle \Psi_0 | {\cal O}
^\dagger_1\, ( 1 + P ) \, G_0 \, | U_1 \rangle ,
\label{eqfad11}
\end{equation}
where ${\cal O}$ is taken as the transverse current operator of Ref.
\cite{bk00}

\noindent
The response function $R_T$ can also be obtained by means of direct 
integrations in Eq.~(\ref{rt}). This gives not only a possibility
for a check of numerics, but provides information about two-- and 
three--body parts of the total cross section. 
%**********************   begin ******************************
\subsection{The Lorentz Integral Transform Method}
%\section{The Lorentz Integral Transform Method}
\noindent
%**********************   end ******************************

\noindent
In the LIT approach the need to compute final state continuum wave 
functions is avoided \cite{ELO94}. In fact the response function $R_T$ is obtained 
via evaluation and subsequent inversion of its LIT

\begin{equation}\label{lorentztr}
{L}(\sigma_R,\sigma_I)=\int_{E_{th}}^\infty 
dE_\gamma {R(E_\gamma)\over(E_\gamma-\sigma_R)^2+\sigma_I^2}\,.
\end{equation}
Neglecting the small recoil term in Eq.~(\ref{rt}) 
$E_\gamma=E_f-E_0$ and applying 
closure the transform ${L}(\sigma)$ of the response $R_T$ is 
found as
\begin{equation}\label{norm} 
{L}(\sigma_R,\sigma_I)=\langle \tilde{\Psi}(\sigma_R,\sigma_I)| 
\tilde{\Psi}(\sigma_R,\sigma_I)
\rangle, 
\end{equation}
where $\tilde{\Psi}$ is the localized solution of the 
Schr\"odinger--like
equation 
\begin{equation}\label{lorentzeq}
({H}-E_0-\sigma_R+ i \sigma_I)\tilde{\Psi}=Q
\end{equation}
with the source term $Q={\bf J}_T\Psi_0$. 
The wave function $\Psi_0$ is the ground state solution 
for the same hamiltonian.

\noindent
In order to calculate $\Psi_0$ and $\tilde{\Psi}$
expansions on correlated hyperspherical harmonics (CHH) are used
\begin{equation}
\Psi\ =\ \tilde\Omega\,\sum_i\,c_i\,\phi_i\,,
\end{equation}
where $\tilde\Omega$ is a correlation operator and
$\phi_i$ is a totally antisymmetric basis set constructed from
a spatial part $\chi_{i,\mu}$ and a spin-isospin part $\theta_{\mu}$ 
\begin{equation}
\phi_i\ =\ \sum_{\mu}\,\chi_{i,\mu} \theta_{\mu}\ .
\end{equation}
The operator $\tilde\Omega$ is a state dependent correlation 
operator. Further details can be found in Ref. \cite{ELOT00}.
Calculating $L(\sigma_R,\sigma_I)$ for a sufficient
number of $\sigma_R$ and fixed $\sigma_I$, one obtains $R_T$ from
the inversion of the LIT (for details of the inversion see Ref.
\cite{ELO99})
%**************************   begin  *******************************
\subsection{Comparison between the two Approaches}
%\section{Comparison between the two Approaches}
%**************************   end   *******************************
\noindent
It is evident that the Faddeev and LIT methods are completely 
different. The Faddeev results are obtained essentially by solving 
Eq.~(\ref{U1}) in momentum space while the LIT results are
obtained by first solving
Eq.~(\ref{lorentzeq}) in configuration space and then inverting 
Eq.~(\ref{lorentztr}) with $L(\sigma)$ given by 
Eq.~(\ref{norm}).
Therefore the comparison between the results
obtained with the two methods is a very significant 
test of accuracy reached by these approaches.

\noindent
There is an interesting similarity between the two
methods, namely, the fact that closure plays an important role
in both approaches. Indeed it is just the use of closure that
leads to Eqs. (\ref{lorentzfad}) and (\ref{lorentzeq}), respectively.
Actually these two equations are equal provided that 
$\sigma_I\rightarrow\epsilon$. However
in the LIT case the finite $\sigma_I$ in Eq.~(\ref{lorentzeq}) 
effectively makes it a bound state problem,
while the Faddeev-like  
equation (\ref{U1}) remains a continuum problem.  

\noindent
In judging the quality of agreement between the results 
of the two methods one has to take into account that the LIT 
results are obtained in the ``unretarded dipole 
approximation''. In the following we recall this approximation
briefly. It consists in replacing
the total current operator ${\bf J}({\bf Q})$  by its limit ${\bf J}(Q=0)$. 
Then one makes use of the continuity equation  
\begin{equation}\label{app}
{\bf Q}\cdot {\bf J}_{f0}({\bf Q})=
E_\gamma \, {\rho}_{f0}({\bf Q}) \,,
\end{equation}
where ${\rho}({\bf Q})$ represents the charge operator.
Expanding ${\rho}_{f0}({\bf Q})$ in powers of ${\bf Q}$ 
and taking into account
that ${\rho}_{f0}(Q=0)=0$, one gets
\begin{equation}\label{bpp}
{\bf Q}\cdot {\bf J}_{f0}(Q=0)= E_\gamma \, 
{\bf Q}\cdot\left[\left(\vec{\nabla}_{\bf Q}\rho\right)_{Q=0}\right]_{f0}.
\end{equation}
As the direction of ${\bf Q}$ is arbitrary, one has
\begin{equation}\label{cpp}
{\bf J}_{f0}(Q=0)=
E_\gamma \left[\left(\vec{\nabla}_{\bf Q}\rho\right)_{Q=0}\right]_{f0}\,.
\end{equation}
Since the gradient of $\rho$ for $Q=0$ is the dipole
operator one writes
\begin{equation}\label{dip}
\left[\left(\vec{\nabla}_{\bf Q}\rho\right)_{Q=0}\right]_{f0}=
i{\bf {D}}_{f0}\,.
\end{equation}
Thus Eq.~(\ref{sigmat}) becomes the well known result for the
photonuclear cross section in unretarded dipole approximation
\begin{equation}
\sigma=4\pi^2 \frac{e^2}{\hbar c}\,E_\gamma\,R_D(E_\gamma)
\end{equation}
with
\begin{equation}\label{crsd}
R_D(E_\gamma)=\int df \,|\langle\Psi_f|{\bf D}|\Psi_0\rangle|^2 \,
              \delta(E_\gamma-E_f+E_0)\,.  
\end{equation}
For the LIT only $R_D$ is calculated. Since our energy range
of interest is below pion threshold the unretarded
dipole approximation should be very good. In the following 
discussion a check of this assumption
will be presented in the framework of the momentum space
Faddeev approach.
In this framework one can use the representation
of the electric multipoles of the transition operator obtained in
Ref. \cite{bk00} employing the continuity equation (\ref{app}) with 
$E_\gamma=Q$:
\begin{eqnarray}
T^{el}_{J\xi}(Q)=T^{\rho}_{J\xi}+T^{res}_{J\xi}\label{elmult} \,,\\
T^{\rho}_{J\xi}=-\frac{1}{4\pi}\sqrt{\frac{J+1}{J}}
\int\, d\hat{Q} \,Y_{J\xi}(\hat{Q}) \, \rho_{f0}({\bf Q})\label{trho} \,,\\
T^{res}_{J\xi}=-\frac{1}{4\pi}\sqrt{\frac{2J+1}{J}} \int\, d\hat{Q}
\left({\bf Y}^\xi_{J,J+1,1}(\hat{Q})\cdot {\bf J}_{f0}({\bf Q}) \right) \,.
\label{tres}
\end{eqnarray}
Here $\xi=\pm1$ and ${\bf Y}^\xi_{J,J+1,1}$ are vector spherical harmonics.
In the limit $Q\rightarrow 0$ the contribution (\ref{tres}) behaves as
$Q^{J+1}$, while the contribution (\ref{trho}) behaves as $Q^{J}$. 
Expressing $T^{\rho}_{J\xi}$ in
terms of the density matrix elements ${\rho}_{f0}({\bf R})$ one finds that
\begin{equation}
\lim_{Q\rightarrow 0}\left[Q^{-J}T^{el}_{JM}(Q)\right]=
\lim_{Q\rightarrow 0}\left[Q^{-J}T^{\rho}_{JM}(Q)\right]
=-i^J\sqrt{\frac{J+1}{J}}\frac{1}{(2J+1)!!}
\int\, d{\bf R}R^JY_{JM}(\hat{R})\rho_{f0}({\bf R}) \,,
\end{equation}
which is known as Siegert's theorem, namely that the electric multipole
transitions are completely determined via the charge density in the limit 
$Q\rightarrow 0$. In particular, for $J=1$ one has
\begin{equation}
\left({D}_M\right)_{f0}=i\sqrt{6\pi}\lim_
{Q\rightarrow 0}
\left[Q^{-1}T^{\rho}_{1M}(Q)\right]\label{lim} \,.
\end{equation}
This is the relation which has been used in the framework of the Faddeev 
calculation to test the quality of the unretarded E1 approximation.

Corrections to the unretarded E1 approximation consist of two parts. 
The first part is due to the difference between $T^{\rho}_{1M}(Q)$ of 
Eq.~(\ref{trho}) and its limit for $Q\rightarrow 0$. We shall refer to 
this correction as a retardation correction. The second part is 
$T^{res}_{J\xi}$ of Eq.~(\ref{tres}).
It is seen from above that both corrections vanish at small $Q$.
Additional contributions come from  magnetic multipoles
(see Eq.~(4) in \cite{bk00}) and higher EJ multipoles.

\section{RESULTS AND DISCUSSION}

For the calculation of the total photoabsorption cross section 
$\sigma_T$ we use the AV18 NN interaction \cite{AV18},  
the 3--body force Urbana IX \cite{UIX} and for the LIT case 
also the Coulomb force. Only the unretarded dipole approximation $R_D$  
is considered within the LIT method. The Faddeev results also include 
retardation effects, additional 
multipoles of the usual one-body current and explicit MEC.

We start the discussion with the benchmark test.
In Fig.~1 the comparison between LIT and Faddeev results 
is presented in unretarded dipole approximation without (Fig.~1a) and
with (Fig.~1b) the 3N force. 
There are two curves for the LIT case showing the small uncertainties 
in the inversion of the transform. 
Besides small differences at higher energies the agreement between the two
methods is excellent for AV18. Inclusion of the 3N force gives slightly poorer agreement in the
peak region and at the low energy side of the peak, where the two curves
have a relative shift of 0.2 MeV. 
We do not think that this small difference is of relevance at 
present; only in case of future high precision experiments 
would it be reconsidered. 
Fig.~1b shows also the result for AV18 only. As already pointed out 
in Refs. \cite{ELOT00,ELOT01} one finds a decrease of the peak and 
enhancement of the tail due to the 3N force.

In Fig.~2 we show the retardation effects (Fig.~2a) and those due
to the additional contributions of $T^{res}_{1\xi}$ and of 
magnetic and higher electric multipoles (Fig.~2b). 
Only the one-body current was employed to calculate these additional
contributions. One sees that the retardation effects reduce the cross 
section. They are very small below 50  MeV (1 \% or less) 
and become somewhat more sizable at higher energies (5 \% at 100 
MeV, 17 \% at pion threshold). As seen in Fig.~2b the increase due to 
additional multipoles is rather similar with and without 3N force. 
Only beyond 80 MeV does
the difference become somewhat more pronounced. At low energy there is a tiny 
decrease due to the additional E1 contribution. Above 30 MeV
the higher multipole contributions are increasing up to 30 \% 
at pion threshold. It is interesting to note that one finds a 
partial compensation of retardation and higher
multipole effects as predicted by Gerasimov via sum rule 
considerations \cite{gerasimov}. 

In Fig.~3 we illustrate the effect of the Coulomb force in the 
threshold region. The neglect of the Coulomb force leads to an increase of 
the cross section close to threshold (about 10 \% at 7 MeV), but becomes 
rather small beyond 15 MeV. Considering the Coulomb force only in the 
ground state is not a good approximation. It is better to neglect 
the Coulomb force completely.

Now we turn to the discussion of MEC effects. Generally it is a problem
to construct a consistent exchange current for a given NN potential. Using,
however, minimal coupling one obtains a MEC with the correct divergence
of the current also in case of realistic NN potentials \cite{Bu85,Ris85},
while the rotor of the current remains model dependent. Here we use only the
standard $\pi$- and $\rho$-like exchange currents for AV18, which are
determined according to the Riska prescription \cite{Ris85}. 
In the present Faddeev
calculation MEC are added to the one--body current without making a 
multipole decomposition, therefore without using Eq.~(\ref{elmult}).
This differs from the usual approach where the gauge condition Eq.~(\ref{app})
is applied to obtain a representation similar to that of Eq.~(\ref{elmult}) 
 and thus only MEC contributions beyond it are calculated 
explicitly. Such a procedure has the advantage that one violates 
gauge invariance in the minimal way since the divergence of the full current
is taken into account correctly. For example, in a calculation with 
3N forces one automatically takes into account a large part of
three--nucleon MEC effects.
Fig.~4 illustrates that the result with explicit MEC differs rather
substantially from that with one--body current and implicit MEC via 
Siegert's theorem. This suggests that the MEC are not fully consistent with 
the potential. Indeed, using Eq.~(\ref{elmult}) for a current satisfying the 
continuity equation the explicit MEC contribution to the E1 multipole 
would manifest itself only in the residual term (\ref{tres}) which vanishes for
$Q\rightarrow 0$. Even away from this limit $T^{res}_{J\xi}$ is
probably  small, as it is in the case of the deuteron \cite{Ar81}.  
In addition, MEC contributions to other multipoles could not decrease the
cross section, since the cross section is the incoherent sum of all 
multipoles and without explicit MEC such multipole contributions are 
negligible at low energies  (see Fig.~2b). 

The contributions of the two- and three--body break up channels are
shown in Fig.~5a.  The three-body break up cross section becomes 
larger than that of the two-body break up already at 14 MeV
and at higher energies it is the dominating channel.
In Fig.~5b we compare the total three--body break up  cross section with 
that of  the final isospin T=3/2 channel (which is three--body break
up exclusively). 
It is interesting to see that at low energy 
the T=1/2 channel gives a small contribution to the three body
break up, while it becomes increasingly important at higher energies.

In Fig.~6 we compare the triton results with experimental data.
One finds a rather good agreement between theory and experiment 
at low energy for both channels. Since the experimental situation 
at higher energy is not settled, a definite comparison between theory 
and experiment cannot be made. In fact the two-body break up data 
are rather scattered and there is limited experimental information 
on the three-body break up.

In Fig.~7 we compare the total $^3$He cross section with data 
(the results for $^3$H with AV18 and UrbIX are presented in
 \cite{ELOT01}). 
One finds a rather good agreement between theoretical and experimental 
data and it is evident that the 3N force improves the comparison with 
experiments at low energy.
On the other hand due to the insufficient precision of the experimental 
data a more conclusive comparison between theory and experiment cannot 
be made.

We summarize our results as follows. Benchmark results of high precision 
have been set for the three--nucleon total photoabsorption cross sections.
The theoretical quality of the results is considerably higher than the
experimental uncertainties. Of course, one has to take into account that
these experimental data are rather old. Thus new experimental activity
in this field would appear
to be timely. The classical approximation for photonuclear cross
sections, i.e. the unretarded dipole approximation, is extremely good 
below 50 MeV. It is also rather good at even higher energies 
(error at 100 MeV below 20 \%), since one has a partial cancellation
of the E1 retardation effects and contributions from additional multipoles. 
The Coulomb force effects on the cross section become negligible about 6-7 MeV 
above threshold but are important at lower energy.

\bigskip
\noindent
Acknowledgement

\noindent
Parts of this work were supported by the Deutsche Forschungsgemeinschaft 
(J.G.), the Polish Committee for Scientific Research (grant 2P03B02818),
a NATO Collaborative Research Grant (W.L.,G.O.,E.L.T.), the National
Science and Engineering Research Council of Canada (E.L.T.),
the Russian Fund for Basic Research, grant 00--15--96590 (V.D.E.),
and the Italian Ministry of Research (V.D.E.,W.L.,G.O.). 
A.N. acknowledges partial support from NSF grant \#PHY0070858.
Parts of the  numerical Faddeev calculations have been performed on the Cray 
T90 and T3E of the NIC in J\"ulich, Germany.

\vfill\eject
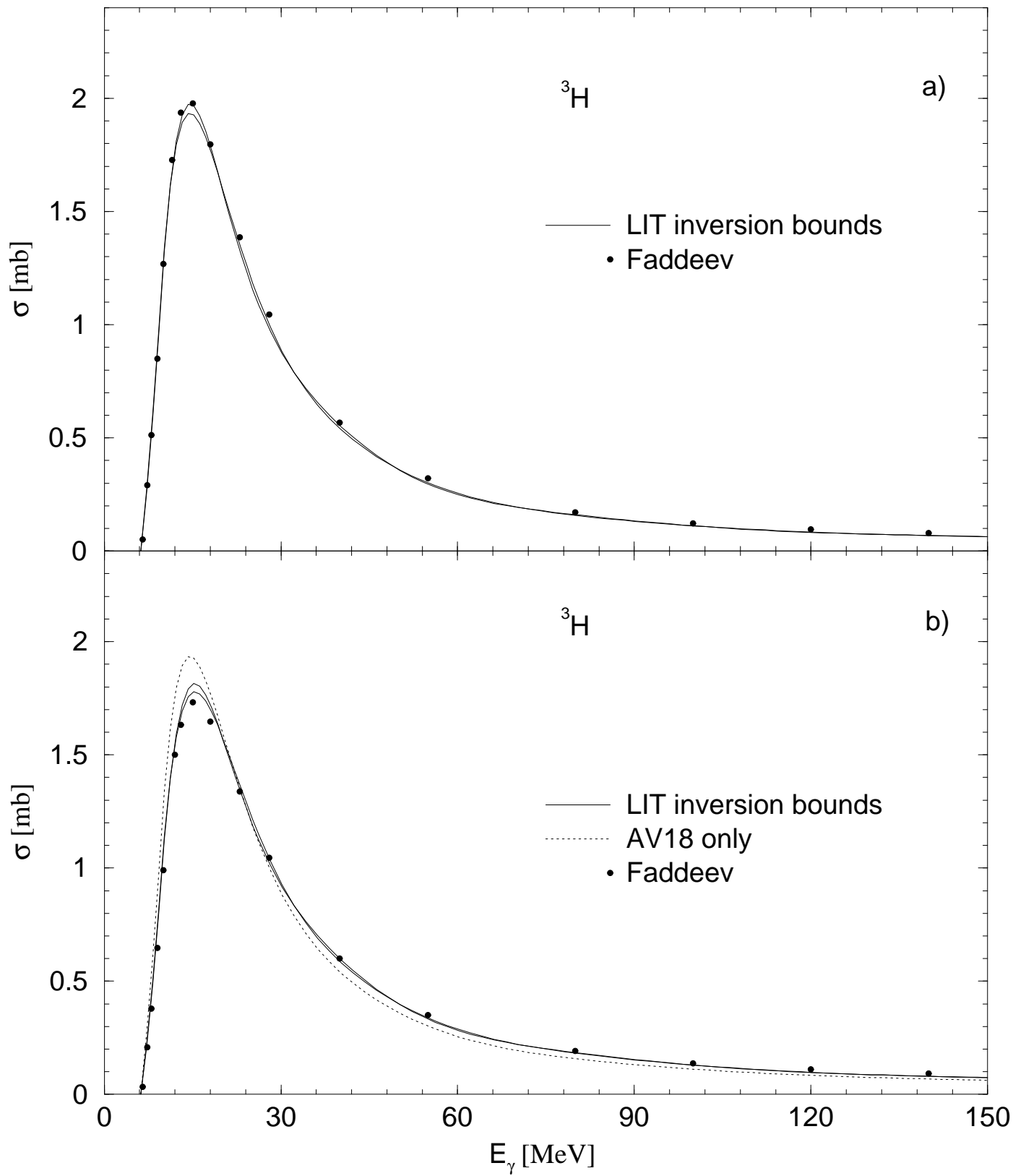
\begin{figure}
\caption{Comparison of the Faddeev and LIT results for the total 
$^3$H photoabsorption cross section in unretarded dipole approximation
without (a) and with (b) 3N force. The dots are the Faddeev results
and the two curves represent the bounds for the inversion of the LIT.
The dotted curve in (b) is the result with AV18 only.}
\end{figure}

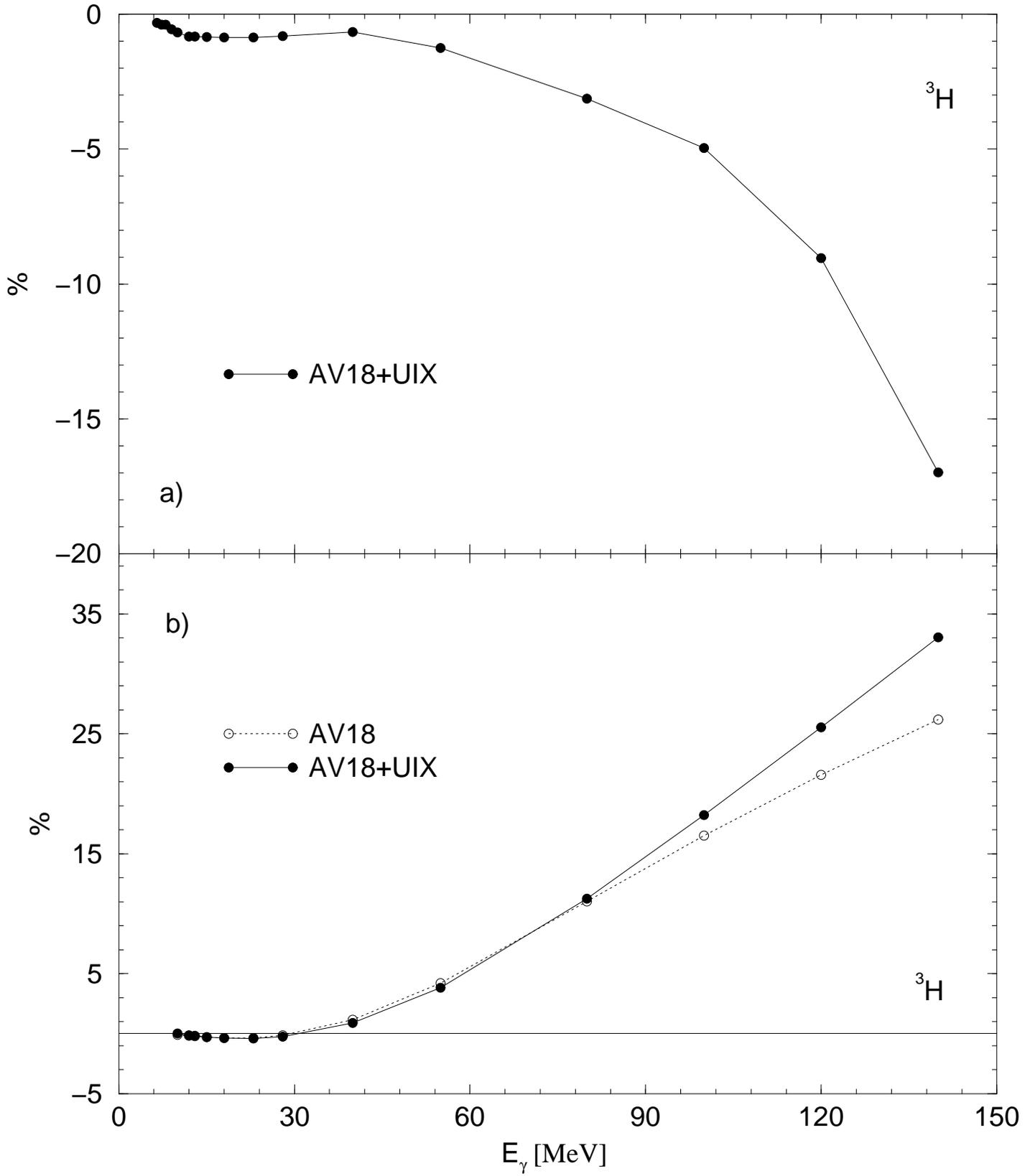
\begin{figure}
\caption{(a) Retardation effects to the unretarded E1 triton cross section 
($\sigma_{ret}- \sigma_{unret})/\sigma_{unret}$ and (b) retardation and
additional contributions (see text)
($\sigma_{tot}-\sigma_{unret})/\sigma_{unret}$.}
\end{figure}

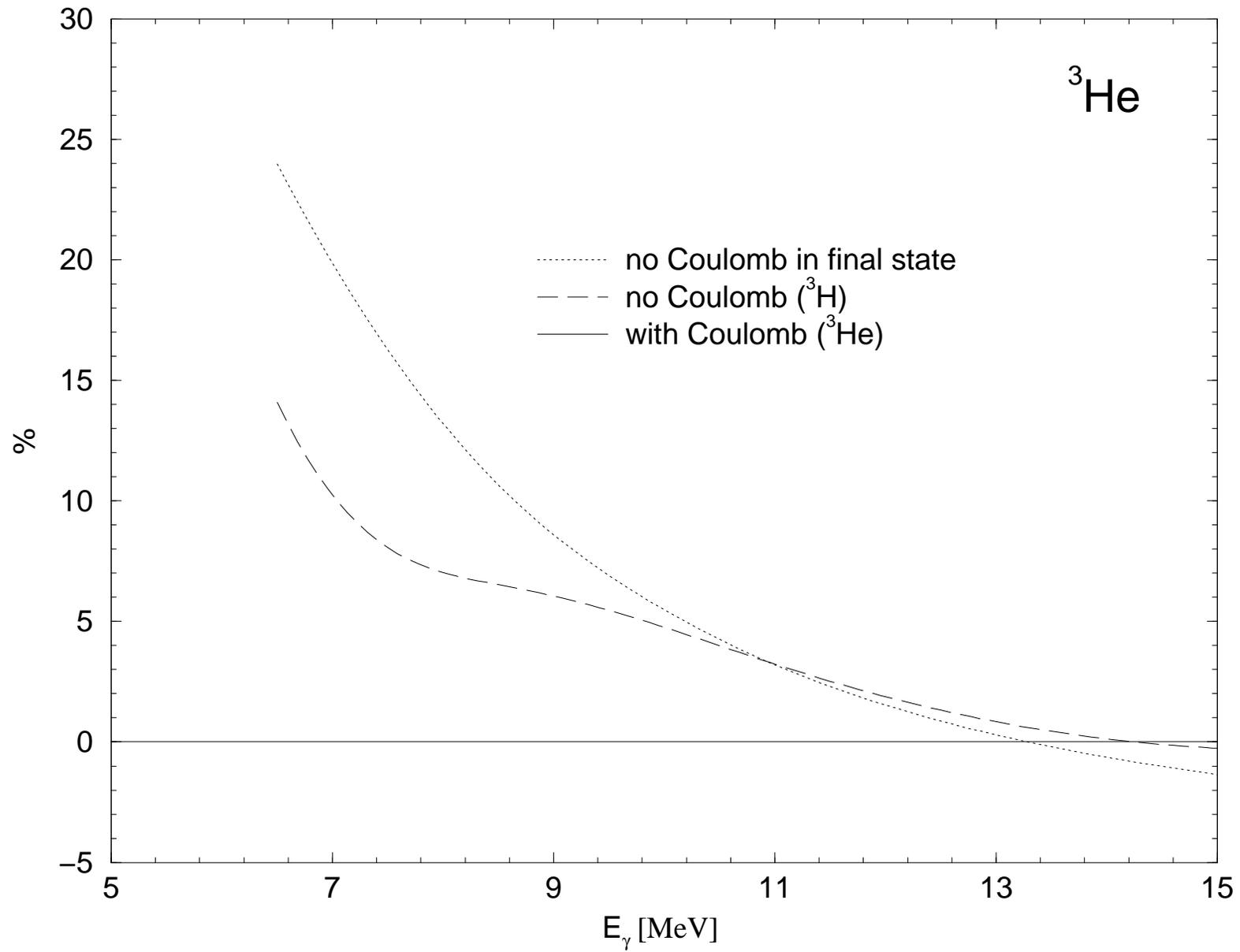
\begin{figure}
\caption{Relative effect of the Coulomb force on the $^3$He  photoabsorption 
cross section: no Coulomb at all, i.e. triton result with shifted threshold
(dashed curve), no Coulomb in final state interaction (dotted curve).}
\end{figure}

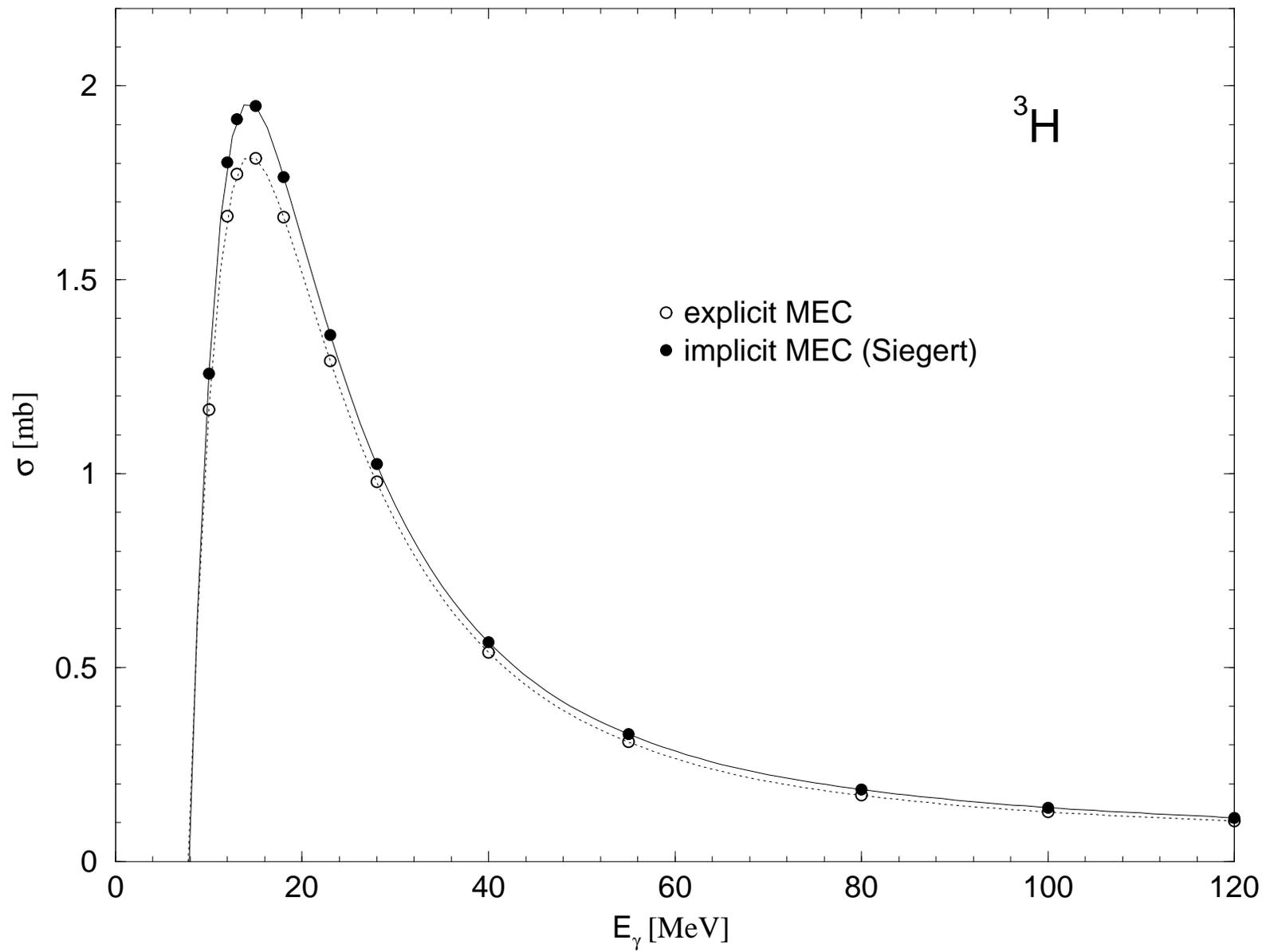
\begin{figure}
\caption{ $^3$He  photoabsorption cross section without a 3N force for the full 
one-body current and implicit MEC contribution via application of Eq.~(24) 
(magnetic one-body transitions considered via Eq.~(4) of Ref. [15]) 
(full dots) and result for the full one-body current with additional explicit 
MEC but without application of Eq.~(24) (empty dots). Curves are spline 
interpolations.}
\end{figure}

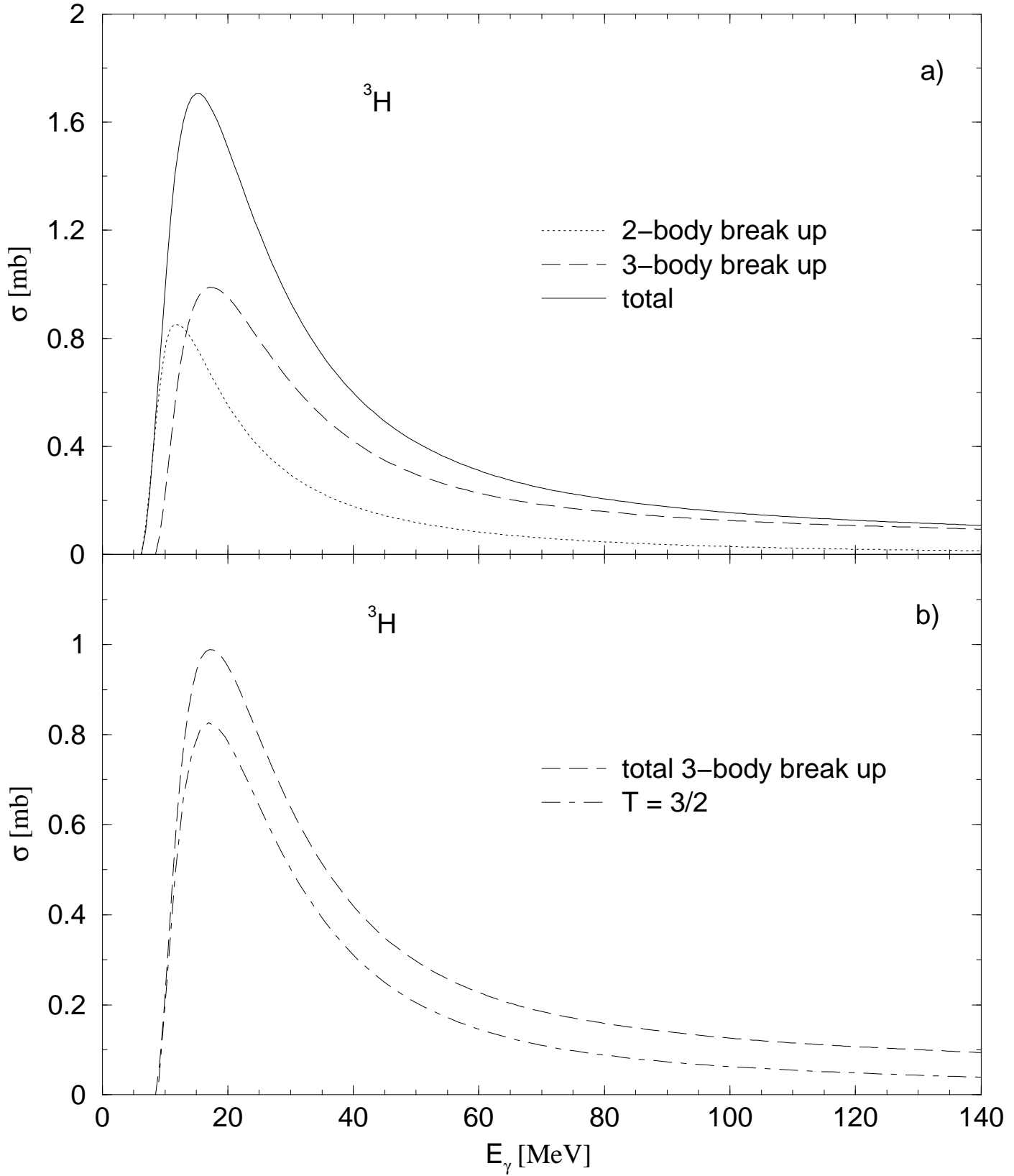
\begin{figure}
\caption{(a) Two--body (dotted curve), three--body (dashed curve) and total 
(full curve) break up 
cross sections of the $^3$H photodisintegration (full one-body current
and implicit MEC contribution as in Fig.~4). 
(b) Comparison of the total three-body break up cross section (dashed curve) 
with the cross section in the final state isospin channel T=3/2 in the 
unretarded dipole limit.}
\end{figure}

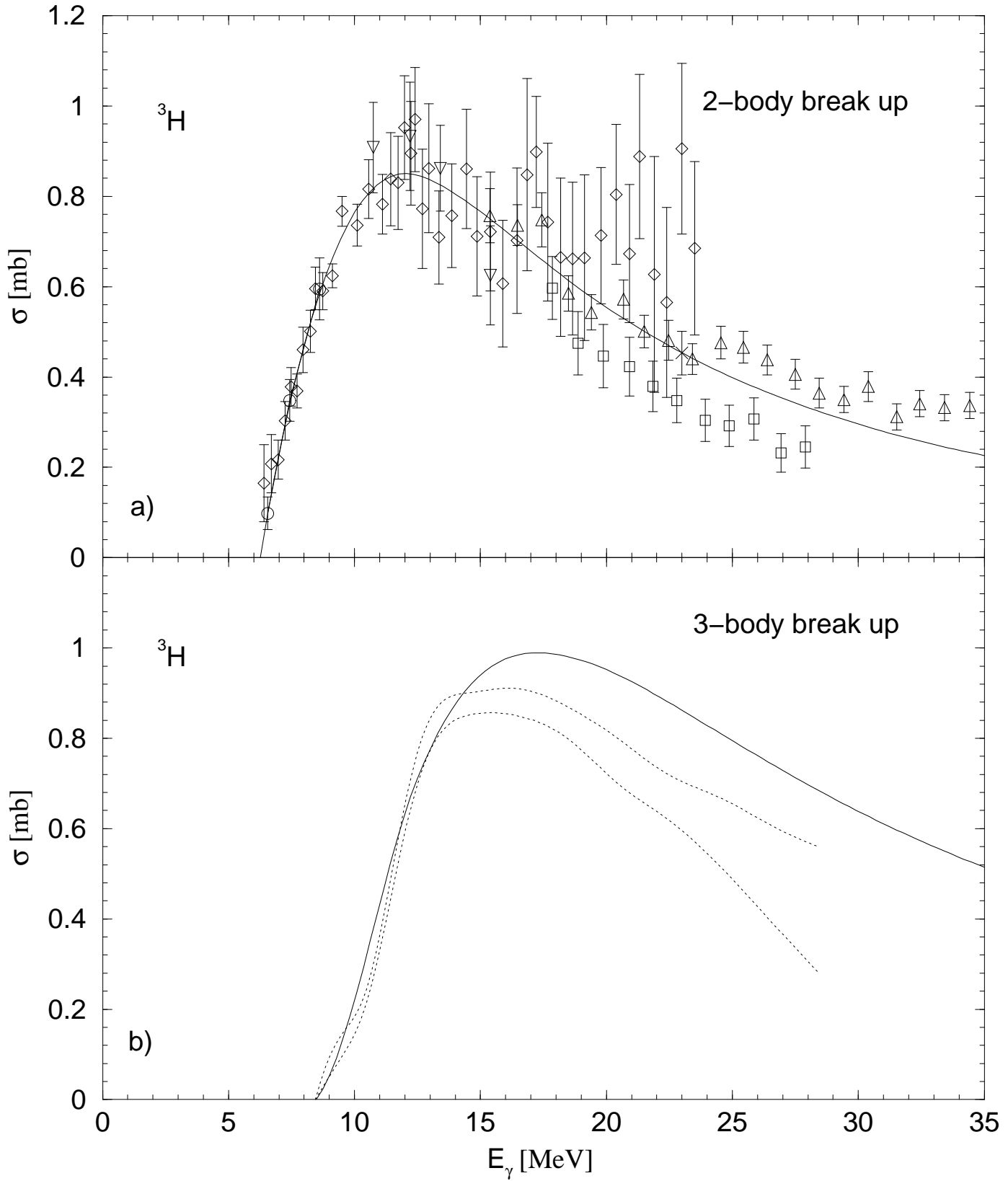
\begin{figure}
\caption{Two--body (a) and three-body (b) break up cross sections for the 
$^3$H photodisintegration (full one-body current and implicit MEC 
contributions as in Fig.~4). The full curves represent Faddeev results. 
Experimental data in (a) from [21] (circles), [22] (squares), [23] (diamonds), 
[24] (triangles up), [25] (triangles down), [26] (X).   
The dotted lines in (b) are bounds of experimental data from [23].}
\end{figure}

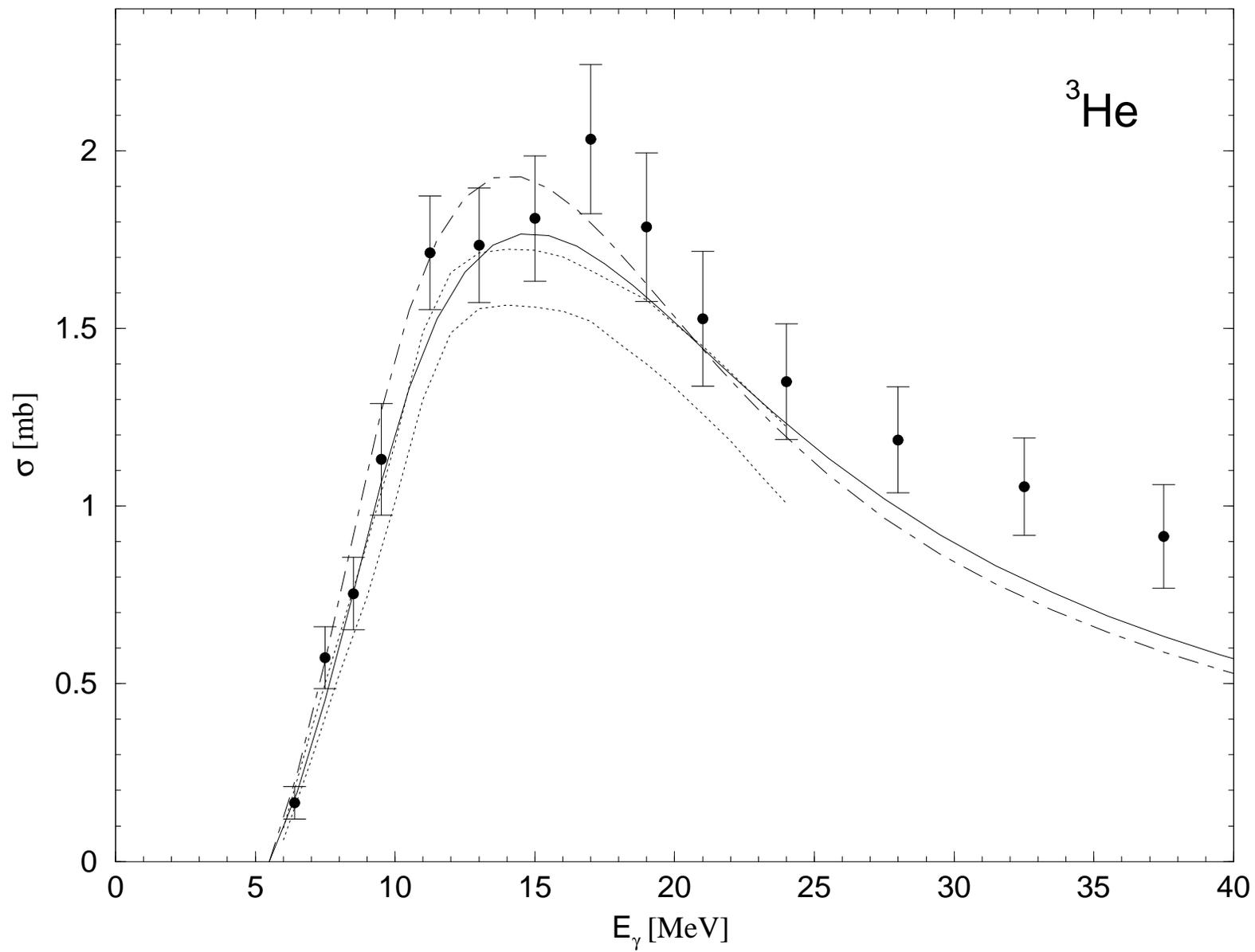
\begin{figure}
\caption{Total $^3$He photoabsorption cross section. LIT results in unretarded 
dipole approximation with AV18 alone (dash-dotted curve) and with AV18+UrbIX 
(full curve); the dotted lines are bounds of experimental data from 
[23] and the dots are data from [27].}
\end{figure}

\begin{thebibliography}{99}

\bibitem{BaP70} I.M. Barbour and A.C. Phillips, Phys. Rev. {\bf C1} (1970) 165.

\bibitem{GiL76} B.F. Gibson and D.R. Lehmann, Phys. Rev. C {\bf 11} (1975) 29; 
{\bf 13} (1976) 477.

\bibitem{Fang78} K.K. Fang, J.S. Levinger, and M. Fabre de la 
Ripelle, Phys. Rev. C {\bf 17} (1978) 24.

\bibitem{Vost81} A.N. Vostrikov and M.V. Zhukov, Yad. Fiz. {\bf 34}
(1981) 344 [Sov. J. Nucl. Phys. {\bf 34}, 196 (1981)].

\bibitem{ELO97} V.D. Efros, W. Leidemann, and G. Orlandini, Phys. 
Lett.{\bf B408} (1997) 1.

\bibitem{ELOT00} V.D. Efros, W. Leidemann, G. Orlandini, and  
E.L. Tomusiak, Phys. Lett. {\bf B484} (2000) 233.

\bibitem{ELOT01} V.D. Efros, W. Leidemann, G. Orlandini, and  
E.L. Tomusiak, Nucl. Phys. {\bf A689} (2001) 421c.

\bibitem{Ar74} H. Arenh\"ovel, W. Fabian, and H.G. Miller, Phys. Lett. 
{\bf B52} (1974) 303.

\bibitem{Bu85} A. Buchmann, W. Leidemann, and H. Arenh\"ovel, Nucl. Phys. 
{\bf A443} (1985) 726.

\bibitem{Ar91} H. Arenh\"ovel and M. Sanzone, Few-Body Syst. Suppl. {\bf 3} 
(1991). 

\bibitem{Gol95} J. Golak, H. Wita{\l}a, H. Kamada, D. H\"uber, 
S. Ishikawa, and W. Gl\"ockle, Phys. Rev. C {\bf 52} (1995) 1216.  

\bibitem{Ishi98} S. Ishikawa, J. Golak, H. Wita{\l}a, H. Kamada, 
W. Gl\"ockle, and D. H\"uber, Phys. Rev. C {\bf 57} (1998) 39. 

\bibitem{ELO94} V.D. Efros, W. Leidemann, and G. Orlandini, Phys. Lett. 
{\bf B 338} (1994) 130.

\bibitem{ELO99} V.D. Efros, W. Leidemann, and G. Orlandini,
Few-Body Syst. {\bf 26} (1999) 251.

\bibitem{bk00} J. Golak et al., Phys. Rev. C {\bf 62} (2000) 054005.

\bibitem{AV18} R.B. Wiringa, V.G.J. Stoks, and R. Schiavilla, 
Phys. Rev. C {\bf 51} (1995) 38.

\bibitem{UIX} R.B. Wiringa, Phys. Rev. C {\bf 43} (1991) 1585; B.S. Pudliner,
V.R. Pandharipande, \\ J. Carlson, C. Steven Pieper, and R.B Wiringa,
Phys. Rev. C {\bf 56} (1997) 1720.

\bibitem{gerasimov} S.B. Gerasimov, Phys. Lett. {\bf 13} (1964) 240.

\bibitem{Ris85} D.O. Riska, Phys. Scr. {\bf 31} (1985) 107; {\bf 31} 
(1985) 471.

\bibitem{Ar81} H. Arenh\"ovel, Z. Phys. A {\bf 302} (1981) 25.

\bibitem{boesch} R.B. B\"osch, J. Lang, R. M\"uller, and W. W\"olfli,
Phys. Lett. {\bf 8} (1964) 120; Helv. Phys. Acta {\bf 38} (1965) 753.

\bibitem{kosiek} R. Kosiek, D. M\"uller, and R. Pfeiffer, 
Phys. Lett. {\bf 21} (1966) 199.

\bibitem{faul81} D.D. Faul, B.L. Berman, P. Meyer, and D. Olson,
Phys. Rev. C {\bf 24} (1981) 849.

\bibitem{skopik} D.M. Skopik, D.H. Beck, J. Asai, and J.J. Murphy II,
Phys. Rev. C {\bf 24}(1981) 1791.

\bibitem{mitev} G. Mitev, P. Colby, N.R. Roberson, and H.R. Weller,
Phys. Rev. C {\bf 34}(1986) 389.

\bibitem{moesner} J. M\"osner, K. M\"oller, W. Pilz, G. Schmidt, and T.
Stiehler, Few-Body Syst. {\bf 1} (1986) 83.

\bibitem{fetisov} V.N. Fetisov, A.N. Gorbunov, and A.T. Varfolomeev,
Nucl. Phys. {\bf A71} (1965) 305.

\end{thebibliography}
\end{document}